# Non-LTE effects on Be and B abundance determinations in cool stars

Dan Kiselman [1], Mats Carlsson [2]

[1]NORDITA, Blegdamsvej 17, DK-2100 Copenhagen, Denmark
[2]Institute of Theoretical Astrophysics, University of Oslo, Box 1029,
Blindern, N-0315 Oslo, Norway

**Abstract:** We discuss the nature of non-LTE effects affecting abundance analysis of cool stars. The departures from LTE of importance for the B I lines in solar-type stars are described and some new results are presented. Boron abundances derived under the LTE assumption have significant systematic errors, especially for metal-poor stars. For beryllium, current results suggest that departures from LTE will not affect abundance analysis significantly.

## Introduction

The interest in stellar abundances of boron and beryllium makes it imperative that the derivation of these abundances from observed spectra is as accurate as possible. The approximation of LTE, local thermodynamic equilibrium, is commonly used in the line-transfer computations needed for abundance analysis. We will discuss the shortcomings of the LTE and the systematic errors it introduces in boron and beryllium abundance determinations. The discussion will be restricted to cool stars, with effective temperatures from 5000 K up to 7000 K.

## What is a non-LTE effect?

The approximation of LTE, local thermodynamic equilibrium, enters twice in the standard procedure of deriving cool star elemental abundances from observations of absorption lines. LTE is first assumed in the computation of a model atmosphere for the star in question. LTE is then used in the treatment of line radiation transfer when calculating equivalent widths or synthetic spectra. The non-LTE effects discussed here are the errors introduced in the second stage. We will discuss the line-formation processes taking place in the standard LTE model atmospheres, but the treatment of the lines of interest is made without assuming LTE .

It should be noted that LTE is not an altogether unambiguous concept – there are several definitions around. Here, we will let LTE stand for the simplifying assumption that the atomic level populations in the stellar atmosphere under study are equal to what they would be in thermodynamic equilibrium given the local values of electron temperature ($T_e$), pressure, and chemical composition. In LTE we thus get the ionisation balance from the Saha equation and the excitation balance from the Boltzmann equation. The emission and absorption properties of the gas, commonly expressed as the absorption and scattering coefficients ($\kappa_\nu$ and $\sigma_\nu$) and the source function ($S_\nu$), can then easily be calculated and the radiative transfer equation solved for the desired frequencies. Note that with this definition of LTE, the total source function will not necessarily be equal to the Planck function since there may be a scattering contribution to the continuous opacities. The connection between these quantities and the emergent light intensity can be grasped from the formal solution of the transfer equation for outgoing vertical rays from a plane-parallel atmosphere

$$I_\nu = \int_0^\infty S_\nu e^{-\tau_\nu} d\tau_\nu,$$

where the optical depth is defined as $d\tau_\nu = (\kappa_\nu + \sigma_\nu)dz$.

The contributions to the total source function and opacity from a spectral line are the line source function ($S_l$) and the line opacity ($\kappa_{l,\nu}$). The line opacity is proportional to the population density of the line's lower atomic level, the line source function is (for the cases of interest here) proportional to the ratio between the population numbers of the upper and the lower levels. In LTE, the line source function is equal to the blackbody value at the frequency of the line: $S_l = B(T_e)$. These facts are useful to keep in mind when trying to understand the results of non-LTE calculations since it is convenient to discuss departures from LTE in the line opacity and in the line source function separately.

The reason that LTE is expected to fail in stellar atmospheres is the strong radiation field which will be important for setting the atomic level populations. Since the radiation is a *non-local* phenomenon we do not expect the population densities to be set by the *local* electron temperature only. In the non-LTE treatment of the line-formation problem we must solve the statistical equilibrium equations that describe the rates with which the atomic states are populated and depopulated. This must be done simultaneously with the solution of the radiative transfer equations.

Even when LTE fails, it is still useful to study the LTE case. It is practical for the abundance analyst to first make an LTE analysis and then apply non-LTE corrections to the "LTE abundances". It is also convenient to describe the line formation processes in terms of deviations from LTE. Thermodynamic equilibrium implies detailed balance, where each rate is exactly balanced by its inverse process. Departures from LTE are departures from detailed balance, caused by a radiation field that departs from the isotropic blackbody radiation characteristic of the local electron temperature. Collisional processes tend to push the system towards detailed balance. LTE is therefore a good approximation if collisional processes dominate in *all* the atomic transitions.

In the recent literature we find accounts of non-LTE mechanisms like *bound-bound pumping*, *ultraviolet overionisation*, *photon suction*, and *photon losses in resonance lines*. Refer to Carlsson et al. (1994) or Bruls et al. (1992) for detailed discussions. Note that these fancy names do not refer to any new basic physical processes, but rather to mechanisms that serve as useful tools for us to understand how the balance produced by the various ionisation, recombination, excitation, and deexcitation processes differs from what the Boltzmann and Saha formulae predict.

## The challenge of non-LTE calculations

Powerful methods now exist for solving the non-LTE line transfer problem in a given atmosphere. We use the operator perturbation technique of Scharmer & Carlsson (1985) as coded in the program MULTI by Carlsson (1986). We employ MULTI version 2.0, in which line blanketing is taken into account in the photoionization, the treatment of background opacities is improved and the many-level treatment is speeded up by using the local operator of Olson et al. (1986).

A major problem in non-LTE work is to assemble a sufficiently complex model atom. Accurate data is needed for many atomic levels, radiative, and collisional transitions, not just for the very line to be studied. Collisional cross sections are especially hard to come by. Large computational efforts aimed at producing high-accuracy atomic data, notably the Opacity Project (see Seaton et al. (1994) and references therein), have made this work easier.

## Neutral boron

The B I lines of interest for cool star abundance analysis are the resonance doublets at 250 nm and at 209 nm. The latter doublet has not yet been extensively observed, but it may be useful for isotopic ratio studies (Johansson et al. 1992).

### Non-LTE mechanisms for B I

Kiselman (1994) investigated the statistical equilibrium of neutral boron in three solar-type atmospheric models and found significant departures from LTE in stars hotter or more metal-poor than the Sun. This section relates to the mechanisms giving rise to non-LTE effects in the metal-poor stellar model investigated in that paper.

The resonance doublets have line source functions that are well described by a two-level atom approximation. This means that the line source functions are determined solely by processes in the line transition: radiative and collisional excitations and deexcitations. Furthermore, these transitions are dominated by the radiative transitions, and, since the lines are weak, their source functions are set by the background radiation field which in its turn is determined by the background opacities. Since the blue and ultraviolet radiation field in solar-type stars generally has a hotter radiation temperature than the local electron temperature we get the effect of *bound-bound pumping* and the line source function becomes greater than the Planckian value: $S_l > B$.

The line opacities are set by the ionisation balance since the overwhelming majority of neutral boron is in the ground state. The departure from the Saha equilibrium can be described as an overionisation caused by two mechanisms working in the same direction. The pumping in the ultraviolet resonance lines increases the ionisation rates since the radiation fields in the ionisation edges of the excited levels are so much richer than in the ground state ionisation edge (at $\lambda$ =150 nm). Furthermore, these radiation fields are stronger than the local Planckian value. Each of these effects is by itself sufficient to cause overionisation.

The pumping and the overionisation will both make absorption lines weaker than in LTE . This non-LTE effect is insignificant for the Sun, but increases in magnitude for hotter or more metal-poor stars. In these stars, the LTE assumption leads to underestimations of the boron abundance. For the well-known halo star HD 140283 ([Fe/H] = $-2.6$), the non-LTE abundance correction amounts to +0.54 dex (Edvardsson et al. 1994).

### New results

We have made non-LTE calculations for a grid of model atmospheres using the B I model atom (31 levels and 114 line transitions) of Kiselman (1994). The model atmosphere code is that used by Edvardsson et al. (1993).

Kiselman (1994) did not include the effect of blending of other lines with the B I resonance lines – the "background" opacities were just the continuous ones. The effect of such blending can be expected to be a moderation of the pumping effect, thus a decrease of the line source function of the observed lines and a damping of the overionisation effect. Both effects will tend to decrease the non-LTE abundance corrections.

A full non-LTE treatment of blended lines of different atomic species is however a formidable task. In our calculations the "background" lines are assumed to be formed in LTE and pure absorption. It is a major task to assemble detailed linelists for the spectral regions of the B I resonance lines. Our treatment of blending is increasingly schematic when going from the 250 nm doublet to the shortest-wavelength resonance lines which are treated in the same way as photoionisations, i.e. the transitions are considered to be fixed with radiation fields computed by the atmospheric code, thus including the effect of line blanketing in an averaged way. It is not our aim to produce detailed synthetic spectra that can be directly compared with observations, but rather to study the departures from LTE and to acquire abundance corrections. The resulting spectra are, however, illustrative, and some are displayed in Fig. 1.

**Fig. 1.** Synthetic spectra of the 250 nm spectral region containing the B I line that has been used for abundance analysis. Spectra are shown for a section of the grid with lg $g = 3.5$ [cgs]. Fulldrawn line: spectrum without boron. Long-dashed line: boron abundance = lg B/H+12 = 0.3. Short-dashed line: boron abundances = 0.6[Fe/H]+2.4. No macro-turbulent, rotational, or instrumental broadening have been applied. The apparent decrease of line strength in cooler stars when the metallicity is increased from 0 to −1 is due to the importance of metals for the continuous opacities – the spectra are normalised on the "continuum" flux

### Non-LTE abundance corrections

The typical abundance analyst needs non-LTE corrections to improve LTE abundance estimates. The ordinary procedure is to calculate both LTE and non-LTE equivalent widths for a range of abundances. The two resulting curves of growth are then used to interpolate the LTE and the non-LTE abundances implied by a certain equivalent width. The non-LTE abundance corrections can be tabulated as functions of LTE abundance or line strength and atmospheric parameters.

There will, however, be complications if the lines are blended as in the boron case. The abundance analyst is here likely to use spectrum synthesis and not just measure equivalent widths. To provide abundance corrections we must then find the way of constructing curves of growth that gives the results most similar to those from the synthetic spectrum fitting procedure.

The definition of equivalent width is

$$W = \int (1 - \mathcal{F}_\lambda/\mathcal{F}_c)d\lambda.$$

When the continuum flux, $\mathcal{F}_c$, is a slowly varying function of wavelength, this definition is unambiguous. But the introduction of blending background lines makes two definitions possible:

A)  $\mathcal{F}_c$ is the flux that would be present in the absence of the studied line. It is then a rapidly changing function of wavelength due to the presence of background lines. This normalisation is physically reasonable since $W$ still measures the fraction of light that is blocked away. It gives, however, different weight to different parts of the measured line according to the variations of $\mathcal{F}_c$.

B)  $\mathcal{F}_c$ is considered to be some constant or slowly varying function of wavelength, e.g. the flux that would be present if only "continuous" opacities were included. $W$ is then equivalent to the line areas measured in typical spectrum plots like the ones in Fig. 1. In that particular case the definition $B$ gives "equivalent widths" that are proportional to the area contained between the full-drawn and the dashed lines. This fact makes definition $B$ seem a practical one, but the physical significance of $W$ is now less clear than with $A$.

Numerical tests show that the difference between the two definitions, as measured by the the difference in curve-of-growth shapes and the difference in non-LTE abundance corrections that follows, becomes significant when the background lines are strong. We consider definition $B$ to be the one that gives results most similar to the abundance analyst's spectrum fitting. The non-LTE abundance corrections presented in figures 2 and 3 are computed that way from our grid. The arrows displayed in the plots show the magnitude and the sign of the non-LTE abundance corrections by connecting LTE abundances (arrow base) and non-LTE abundances (arrow tip). Each panel shows the corrections for a range of boron abundances and metallicities. The four panels make it possible to see how the exclusion of background lines, increase of effective temperature, and increase of surface gravity affects the size of the corrections. It is clear that the corrections are most important for metal-poor stars and that they increase in size with increasing temperature and decreasing gravity. As expected, the importance of the background lines for the results are greatest in relatively metal-rich stars where they essentially remove the departures from LTE. It is, however, likely that our treatment of the background lines as formed in LTE and pure absorption is least accurate when these lines are strong. The effect of *photon escape* which will depress $S_l$ and tend to strengthen the line relative to LTE is probably underestimated in those cases.

The plot in Fig. 4 shows the results of application of these results to literature LTE abundances of metal-poor solar-type stars.

**Fig. 2.** Non-LTE corrections to boron abundances. Each arrow connects an LTE and a non-LTE abundance for one specific equivalent width. The spread in the position of the arrows for each metallicity ($-3$, $-2$, $-1$, and 0) has been introduced for clarity. The upper panel shows the correction when "background" lines are included in the calculations, in the lower panel these lines are excluded

**Fig. 3.** Non-LTE corrections to boron abundances, similar to Fig. 2. Comparison to the upper panel of Fig. 2 shows how the non-LTE corrections vary with changing effective temperature and surface gravity

**Fig. 4.** Boron abundance as function of metallicity. The LTE results (stars) have been taken from Edvardsson et al. (1994) for HD140283 (big star) and from Duncan et al. (1992) (small stars). The non-LTE abundance corrections have been interpolated from the grid presented in this paper. Big diamonds are results with background line blanketing included, small diamonds without. The solar boron abundance is from Kohl et al. (1977). Indications of beryllium abundances are adopted from Duncan et al. (1992)

## Beryllium

The spectral feature of interest for cool star abundance analysis is the Be II resonance doublet at 313 nm, just longwards of the atmospheric transmission cutoff. We have compiled an atomic model representing Be I and Be II (with 71 levels and 243 lines) and investigated the non-LTE formation of the beryllium lines in three atmospheric models representing the Sun, Procyon, and HD140283. The results should be considered as more preliminary than the boron results discussed above, since the collisional data still needs to be improved. For reasons to be discussed, however, it seems likely that the general results presented here will survive coming improvements.

For the HD140283 model, it seems that same general mechanisms as in the boron case just discussed cause the departures from LTE in beryllium. There is overionisation, apparently caused in the same way as in boron: bound-bound pumping in ultraviolet resonance lines plus a hotter-than-local-Planckian radiation field in the important ionisation edges. There is also pumping in the 313-nm doublet, causing its line source function to increase above the local Planckian value. The difference relative to the boron case is that the interesting lines now are lines arising from the ion: Be II This means that the overionisation will tend to make the lines stronger while the bound-bound pumping will weaken the lines. These effects effectively cancel each other. The lines *cannot* be said to be formed in LTE, but their equivalent widths do not differ much from the LTE -values! The resulting abundance correction is just +0.03 dex.

Since both the overionisation and the pumping effects are driven by the same phenomenon – the hot radiation field in the UV – they can be expected to cancel each other also if this should change by improvements in model atmospheres or background opacities. The non-LTE corrections for the Sun and Procyon are indeed also very small, less than 0.1 dex in both cases. It seems likely that non-LTE effects are not very significant for the 313-nm lines in solar-type stars. The largest effects are expected for very beryllium rich stars (if such exist) when the 313 nm lines are strong and affected by photon losses. The abundance corrections will in that case be negative.

The results for the Sun, in the sense of the line-strength in LTE being similar to non-LTE, is in agreement with earlier work by Chmielevski et al. (1975) and Shipman & Auer (1979).

## Conclusions

Non-LTE effects should be taken into account when deriving abundances from B I lines in cool stars. So far, it seems that the assumption of LTE will not introduce significant errors in the derivation of Be abundances from Be II 313 nm in solar-type stars.

## References

Bruls J.H.M.J., Rutten R.J., Shchukina N.G., 1992, A&A, 265, 237
Carlsson M., 1986, Uppsala Astronomical Observatory Report No. 33
Carlsson M., Rutten R.J., Bruls J.H.M.J., Shchukina N.G., 1994, AA, in press
Chmielewski Y., Müller E.A., Brault J.W., 1975, A&A, 42, 37
Duncan D.K., Lambert D.L., Lemke M., 1992, ApJ, 401, 584
Edvardsson B., Andersen J., Gustafsson B., et al., 1993, A&A, 275, 101
Edvardsson B., Gustafsson B., Johansson S.G., et al., 1994, A&A, in press
Johansson S.G., Litzén U., Kasten J., Kock M., 1993, ApJ, 403, L25
Kiselman, D., 1994, A&A, 286, 169
Kohl J.L., Parkinson W.H., Withbroe G.L., 1977, ApJ, 212, L101
Lemke M., Lambert D.L., Edvardsson B., 1993, PASP, 105, 468
Olson, G.L., Auer, L.H., Buchler, J.R.,1986, JQSRT, 35, 431
Scharmer G.B., Carlsson, M., 1985, J.Comput.Phys. 59, 56
Seaton M.J., Yu Yan, Mihalas D., Pradhan A.K., 1994, MNRAS, 266, 805
Shipman H.L., Auer L.H., 1979, ApJ, 84, 1756